# For a New Department of Energy lab to examine laser fusion for energy

Wallace Manheimer
Retired from the US Naval Research Laboratory

Abstract:

This paper gives a summary of a talk by the author at the mini-conference entitled *Progress in Making IFE-based Concepts a Reality* at the APS-DPP meeting in Atlanta in October, 2024. It argues principally for a new DoE lab to examine the potential opportunity of laser fusion for civilian energy, by direct drive, with an excimer laser. This work is motivated mostly by the demonstration of a burning plasma in an indirect drive configuration, by the Lawrence Livermore National laboratory with its NIF laser. Also, it briefly gives some impressions of the mini conference.

Key words: laser fusion, fusion breeding, excimer lasers

1. Introduction

On the Tuesday, October 8, 2024, of the recent American Physical Society Division of Plasma Physics (APS-DPP) conference in Atlanta, Dr. Kruse of the Lawrence Livermore National Laboratory (LLNL) hosted a mini conference called *Progress in making IFE* (Inertial Fusion Energy) *based concepts a reality*. It was mostly a conference of scientists participating in a new Department of Energy (DoE) effort on laser fusion. This author is not a participant in this effort, but was invited to the mini conference, nevertheless, as he had done work in the area. I gave a talk at it entitled: *Fusion, it is time to color outside the lines* (1), which is the title of a paper making these points in much more detail (2). I also made many of the same points in a book, published by the Generis publishing company entitled: *Mass delusions, how thy harm sustainable energy, climate policy, fusion and fusion breeding* (3). Reference 2 is backed up by 117 references, the work of hundreds of scientists at many different labs, universities and companies. Sections 4 and 5 of Reference 3, the portion dedicated to fusion, is backed up by 87, so this paper hardly relies on the author's work alone, as one might think from the few references here. The real reference list is that in (2 and 3). The reference list of this manuscript list is kept brief only for considerations of the length of this paper, which is intended to be rather short, certainly compared to (2). Since (2 and 3) are already in the literature, with (2) published open access, I will be brief in my discussion of it here.

2. The main point; an alpha generated fusion burn

The main point was that the NIF laser at the (LLNL) achieved a burning plasma about 20 years before ITER hopes to do, and at much lower cost. The laser beam is a match, like the spark plug in the cylinder of a car. It does not burn the fuel, it only ignites it. The alpha fusion burn is analogous. The laser compresses and heats a small part of the target until it starts a tiny fusion reaction, generating 14 MeV neutrons, and 3.5 MeV alpha particles. The geometry of the target is such that the neutrons escape, but the alpha particles are reabsorbed locally, heating the nearby parts of the target to fusion conditions, initiating a burn wave. In a typical explosion, the expanding gas cools as it expands, as thermal energy is converted to kinetic energy. However, LLNL has diagnostics showing that as the exploding target expands, for a while it heats, until ultimately as the expansion proceeds it then cools.

3. How to optimize this amazing accomplishment

To this author, this is an inflection point, which means that American Department of Energy (DoE) fusion effort cannot simply proceed with business as usual but must consider carefully alternate strategies. Specifically, the American DoE should set up a new National lab with the goal of examining and exploiting this amazing development. Its goal should be to use it to develop sustainable energy for the civilian economy. Before this, ITER seemed to be the only reasonable path, but LLNL's amazing achievement changed this completely. The talk asserted that 100 years from now, it could well be regarded as one of the main experiments of the $21^{st}$ century, and that it is Nobel Prize worthy. It seems that because of this triumph, or as just described, this inflection point, the US Department of Energy should shift its main attention from magnetic fusion energy (MFE) to IFE. In the fusion program, there is a precedent for such an abrupt shift. About 60 years ago, the Princeton Plasma Physics Laboratory (PPPL), was working on stellarators, which had very poor confinement. When they learned of the advances the Russians made with tokamaks, nearly immediately, they jumped ship and put all their efforts in tokamaks. This action was amazingly successfully on their part. From about 1970 to about 2005, PPPL led the world in fusion.

References (1 and 2) made the point, and make it again here, that the US DoE should shift its main effort to laser fusion, based on the LLNL results. This is just what the Princeton lab did about 60 years ago but based on Russian tokamak results. Of course, as pointed out in (2), I have no expectation that the grand pooh bah's at the DoE will read this, or (1) or (2), and immediately set up the new DoE lab. Hopefully this might help to initiate discussion among many interested parties and help to convince them that this is an optimum way for the US fusion program to proceed. Perhaps this, as well as (1,2 and 3) can be an initial value in what hopefully will become an exponentially increasing interest in the concept. This could well be the best way for the United States to recover its long-lost leadership of the entire fusion program. After all, if MFE proves to be the way to go, the rest of the world will do it. But we are the only ones able to do laser fusion at this point, and at least to this author, laser fusion now seems to have all the advantages over MFE. I briefly discussed this in my talk (1) and documented it much more thoroughly in (2 and 3).

4. Other considerations

The laser fusion effort at LLNL is not supported by fusion energy, but by DoE NNSA which is concerned with nuclear weapons which are generated by X-ray driven implosions (i.e. stockpile stewardship). The LLNL experiment uses what is called an indirect drive configuration. The target is enclosed in a small container, called a hohlraum, made up of heavy metals, typically gold or uranium. However, the target does not touch the walls of the hohlraum, it is supported in the hohlraum by other means. The laser focuses on the inner walls of the hohlraum creating a blackbody plasma with a temperature of 250-300 eV. The X-rays emitted by this black body illuminate, compress and heat the target. NNSA's interest is only in X-ray driven implosions, and has little or no interest ultraviolet driven implosions, laser efficiency, laser average power, laser bandwidth, rep-rated pulse power, tracking and engaging a fast-moving target in a not so gentile environment, manufacturing targets cheaply in mass…… All of these are vital for laser fusion for civilian energy, and they are all serious challenges. However, to this author, they seem easier to overcome, as they are more technical, than do the more fundamental challenges confronting MFE (1-3). Accordingly, this author has suggested that the proper course of action for DoE is to set up a completely different laboratory to examine and optimize recent results for energy rather than nuclear simulation. This new lab should not be LLNL, as the goals and scientific challenges of the weapons program and the civilian power program are so different; the labs should be separate.

The talk (1), and (2 and 3) in much greater detail made a few other points. Hohlraums now cost many thousands of dollars now and contain very expensive materials like gold and uranium. Surely mass manufacturing can bring their price down but consider indirect drive for the civilian sector. Say we use a 1 MJ laser pulse and find a gain of 100, giving 100 MJ of fusion energy. Converting it to electricity gives about 30 MJ, or ~ 10 kWhrs, worth about a dollar! Can the price of the hohlraum and whatever supporting material it needs be reduced to less than about a dime, or even quarter? It seems like a stretch. What can you get for a dime these days?

For energy, rather than nuclear weapon simulation, direct drive seems to be the way to go. Hence this new lab should concentrate on direct drive, as its goal will be energy for the civilian sector. In a direct drive configuration, the laser is focused directly on the spherical target, and it is imploded by ultraviolet light, rather than by X-rays. Also, with indirect drive, every shot destroys something pricy, the hohlraum, to produce something very cheap, a few kWhrs of electric energy. Another problem is that as the hohlraum and target is shot in, surely on a slightly uncertain, wobbling path, its position *and* its orientation must be exactly aligned with the laser light. This is a not nearly as serious an issue with direct drive and spherical targets, where there is no concern with target orientation. Furthermore, as LLNL readily admits (2 ref 94, 3 ref 55 in Section 4), only about 10-15% of the laser light is absorbed by the target (the target absorbs the X-rays produced on the hohlraum wall). Hence, since their maximum gain, i.e. Q, at the time of this writing is ~2.5 (4), their Q for direct drive, assuming that ultraviolet works as well as X-rays may be as high as 25! Also, for a rep rated system for energy, the reaction chamber must be prepared between shots. In direct drive, one only must clean up the residue of the target; indirect drive, it must clean up the residue of the target and the much larger hohlraum. But while this new lab will focus on direct drive, of course LLNL will continue with its indirect drive, NNSA supported project. Possibly their approach will ultimately be the best way to go, and this new lab should follow its work carefully (and of course visa versa).

While the LLNL result is certainly the tallest poll of the tent for reorganization of the fusion effort around laser fusion, there are important supporting results also from two other labs. First, the Naval Research Lab (NRL) has decades of experience of developing excimer lasers, both KrF and ArF (2, see Ref 87). These definitely have advantages over glass in that they have shorter wavelength and zooming capability. They most likely also have an advantage over glass of higher efficiency and higher average power capability, since they use a flowing gas, and there is no glass to cool down between shots. Other laser types, for instance a diode-pumped, solid-state pump laser, like LLNL's Mercury project might be in the running, and this was discussed in the NRL/LLNL led HAPL program (2, see reference 100 there). However, this author believes that excimer lasers are the way to go.

Since both nuclear weapons and the LLNL burning plasma are driven by X-ray implosions, some have argued that this is the only viable approach to laser fusion. However, the University of Rochester Laboratory for Laser Energetics (URLLE) has produced what they call a hydrodynamically equivalent implosion with their 30 kJ ultraviolet OMEGA laser (2, see Refs 97 and 98). This is an implosion, just like a Megajoule X-ray driven implosion, but with shorter scale length, shorter time, and obviously achievable with much less energy. Hence it may be that the URLLE direct drive experiment makes a convincing argument that an ultraviolet driven implosion could rather quickly reach a Q of 25! To this author, the LLNL burning plasma result, enhanced by the NRL work on excimer lasers, and by the URLLE results on hydrodynamically equivalent ultraviolet driven implosions; strongly emphasize the case that the Department of Energy should support a new lab, based on direct drive and excimer lasers, to develop laser fusion for civilian power.

Another point I made in my talk is that the DoE should no longer persist in its decades long policy of ignoring fusion breeding, that is using fusion neutrons to breed fuel for thermal nuclear reactors. Assuming the world continues to build many thermal nuclear reactors, availability of mined uranium will become a serious issue in the next few decades. Fusion breeders can fuel these reactors, fission breeders cannot, as described in (1-3 and other references therein). Basically, it takes 2 fission breeders, at maximum breeding rate, to fuel a single thermal reactor of equal power. But a single fusion breeder can fuel 5 or 10. Also, since the demands on the pure fusion reactor are much greater than those on the fusion breeding reactor, fusion breeding provides an insurance policy if the calculations of fusion gain, prove to be optimistic, as they nearly always have. For instance, a steady state (or high duty cycle) tokamak like ITER, assuming it is successful, could be fine as a breeder (2, see refs 50-57, and 3); that is ITER could be an end itself. However for pure fusion, according to the ITER web site 'THE DEMO' is what will be needed for commercial fusion. This DEMO must have higher gain, higher power, and yet be smaller, and cheaper. Hence ITER is only an initial steppingstone to who knows what DEMO, at who knows what cost, and after who knows how many decades, assuming the DEMO can be accomplished at all.

For the laser fusion case, it is argued in (2) that a gain of 50 with a 7% efficient laser could be viable for fusion breeding. The most recent LLNL results, interpreted as they might be (optimistically) for direct drive, could already be halfway there! Second, even if the IFE gain optimistic estimates prove to be correct, fusion breeding can greatly lower the cost of fusion provided electricity, even neglecting the additional fuel it provides for thermal reactors (2). Fuel for these thermal reactors could finally become 'too cheap to meter!'.

Finally, (2) had a section on digressions. These are brief, preliminary discussions of matters that could be important. There, a new type of cylindrical target chamber which could have many advantages was introduced. It would have the ability to use as a blanket one or more flowing liquids with free surfaces; or liquids flowing in pipes; or a solid blanket, which could be removed and reinserted very easily. It also pointed out that if fusion, or fusion breeding, does reach a point where it would be necessary to build many fusion reactors quickly, availability of tritium, even for the first commercial reactor, will be a serious problem. The digression in (2) suggested a way to solve this problem of tritium for the first one or two commercial reactors using the world's thermal reactors and then using a reaction in the fusion blanket which could allow exponential growth of tritium supply. Note that it is likely that in the life of a fusion economy, there could be at least 3 different blankets that the reactor might need to use. First there is a blanket for the exponential growth of the tritium supply. Second there is a blanket for pure fusion, where there are already enough fusion reactors, and the extra growth of tritium is the last thing anyone wants (tritium is an important component of the most powerful nuclear weapons). Third there is a blanket for fusion breeding, where some thorium is mixed in. A point source for the fusion reaction, enhanced by the cylindrical blanket just mentioned is likely the only way that a fusion blanket, especially a solid one, can be rapidly changed as needs change. It is very, very difficult to see how a tokamak or stellarator can support these multiple blankets on any kind of reasonable time scale.

5. It is unlikely that the privately funded fusion start-up will deliver power to the grid anytime soon

Also discussed in (2 and 3) is the author's assertion that the new (and not so new) privately funded 'fusion start ups', which promise fusion for the grid in about a decade, are extremely unlikely to deliver on those promises. In (2) there are 10 separate citations where a variety of fusion experts, including the author, mostly retired and financially independent, debunk the claims of fusion to the grid in a decade. In fact, the probability that these 'startups' will supply any net power to the grid in the next decade or so, is about the same as the probability that the NY Mets will put me in right field instead of Juan Soto.

Googling "How long does it take to build a 1 GWe nuclear power plant?", one first sees the AI assessment. Here it is:

Building a 1 GW nuclear power plant typically takes between 5 and 10 years on average, depending on factors like location, regulatory environment, and design complexity, with some countries like South Korea and China potentially building them faster than others; however, some projects can take significantly longer due to delays and complications.

Heck, here is AI's assessment of how long it takes to build a 1 GWe coal fired powered plant:

Building a 1 GW coal-fired power plant typically takes between 3 to 7 years to complete, depending on factors like location, construction complexity, and regulatory processes; with most estimates falling within a 4-5 year timeframe.

These are the times it takes to build power plants where the science and technology are well known! Is it credible that a fusion pilot plant, where the science and technology are far from known, can be built and hitched up to the grid as fast? This author emphatically believes that the answer is NO!

But some fusion 'startups' think they can do it even faster. For instance [Helion Energy](#) has contracted to sell Microsoft 50MW of electric power in 2028! (50 MW for a microsecond perhaps?) However, unlike many of the recent skeptics of the potential of fusion, this author has never wavered in his confidence that fusion, and/or, fusion breeding, is likely to not only be successful, but might well be the salvation of future civilization.

This author is especially worried that these privately funded fusion start-ups could ultimately harm the more normal government supported efforts, despite the billions pouring into them. They are making outrageous claims of fusion to the grid in less than a decade. Utilities using coal can hardly make this claim! When these efforts all collapse, with gigantic thud, as they almost certainly will, who knows how much the already weak credibility of the fusion effort will be damaged. Could they take down with them the more reasonable government sponsored efforts? Very possibly!

6. Ways to fund this new lab

Hopefully new support can be found for this new laboratory. This would certainly be justified by the recent results just mentioned. However, considering that the US government budget deficit is now in the trillions, it may not be possible. If not, the support should come from switching a major part of the MFE support to laser fusion. After all, no federal research program is guaranteed eternal life. As knowledge advances, and as needs change, the supported federal research programs must change with it. Currently, MFE is supported at ~ $500M per year and in addition there is ~$200M per year for the US support of ITER. As ITER is an important international project, which may prove yet that the tokamak approach is viable for commercial fusion, or fusion breeding; this support should remain, especially as ITER is now mostly built. However, pulling a number from the air, about $350M per year should be switched to support this new lab. In providing this support, it should recognize that getting commercial fusion power is certainly a multi decade project; nobody will have it in the next 5 or 10 years. After all, NIF was approved in 1995 for

$1.1B, to be completed in 2002. It was finally completed in 2009, for ~$3.5B! How can the much larger project of setting up an appropriate rep rated laser and pilot power plant possibly be done more quickly and cheaply?

On September 26, 2023, I wrote to the president of Princeton University, the ultimate boss of its PPPL lab and suggested that he propose this change to DoE, just as his predecessors did 60 years ago. Princeton has a nearly unique opportunity here. It not only has a great deal of expertise and infrastructure in fusion, but it also has an endowment of ~$40B and could easily sweeten the pot for DoE by shaking loose some $100-200M. This sounds like a lot of money, but it is much less than the daily fluctuations, in the various markets, of the value of Princeton's endowment. This would provide a rep rated laser with perhaps 100-200 kJ shots (an intermediate energy before the final multi Megajoule laser is built), some new infrastructure, and some new hires for the lab. With this laser it would start to do rep-rated experiments on tracking and hitting fast moving spherical targets, among other tasks. For this *small* investment Princeton might be able to persuade the DoE to choose it for a new lab, and in doing so recover its position as the world leader in fusion. This sounds worth it to me. While I am unwilling to circulate the letter, I did give a brief description of it in the open literature (5).

Certainly, other labs, LANL, ORNL…. could also make their cases. It is also worth mentioning URLLE. It has done a great deal of work in laser fusion, but it is not a national lab, and it has used only glass lasers and has never investigated excimer lasers. The question is whether it could become a DoE national lab and separate from the University of Rochester in the same sense as PPPL has separated from Princeton University. Of course, the University of Rochester has an endowment which is tiny compared to Princeton's. In any case, issues like these are far above my pay grade.

7. Other observation from the mini conference

Now I will discuss my impressions of several other talks at the meeting. The first 3 talks were given first by Dustin Froula of URLLE (6), the leader of a consortium on laser plasma instability (LPI) research; the second was by Carmen Menoni of Colorado state (7), the leader of a university consortium on laser issues; the third by Tammy Ma of LLNL (8) on a coordinated national plan.  What struck me about these talks is that the DoE does not seem take this effort seriously.  Compared what is needed, the support for these projects is chicken feed.  Consider Carmen Menoni's consortium.  I spoke to her and learned her support was ~$16M over four years, split among 5 universities.  I forget Dustin Froula's numbers, but they were comparable.  Yet each had viewgraphs showing far too many tasks for any of us to absorb, and a timeline leading to a fusion pilot power plant in 2035.  Considering that it took 14 years just to build NIF, another 10 to get a burning plasma, all costing at least two orders of magnitude more than what is available in these consortia; the plan is nothing if not extremely optimistic.  This author feels that what the DoE needs to do is to rock the boat and quickly set up a crash program.  Increasing the budget by a million or two every year, so as not to ruffle too many feathers, will at best, delay the development of laser fusion by decades, and at worst kill it completely. That is the author's motivation for his recommendation that the DoE set up a new lab as quickly as possible.

This author finds it difficult to keep the following thought out of his mind: namely that the main part of the DoE is offering table scraps just to quiet down some pesky nuisances, without rocking the boat.  However, to make "*IFE based concepts a reality*", we do not need table scraps, we need the entire banquet.  In other words, several hundred millions of dollars for yearly support for a new lab, for a period of decades is what is required.

Furthermore, even if the DoE came up with the necessary support, which I approximated very roughly a few paragraphs ago, is it really a good idea to have the support split among a couple of dozen completely independent organizations and tell them to work together?  Isn't it better to have a single organization responsible for managing the entire project, like NNSA does with LLNL?  If we have learned anything from ITER, with all its delays and cost overruns, it must be that having 7 different independent leaders, each with its own culture and manufacturing strategy is a serious mistake.  Instead of having the IFE wagon pulled by 100 cats, let's get a horse.  This horse is the new DoE lab proposed here.

8.  Conclusions

If the USA wishes to reestablish its long lost lead in fusion research, this paper makes the case that the best way for it to do so, is by setting up a separate DoE lab to investigate laser fusion using an excimer laser and a direct drive configuration.

Acknowledgement

This paper was not supported by any organization, public or private. I appreciate Dr. Kruse for organizing this meeting and inviting me.

RISE -Inertial fusion science and technology hub- Goals to advance Inertial Fusion Energy, APS DPP Meeting, Atlanta, October, 2024, https://meetings.aps.org/Meeting/DPP24/Session/JM10.4

8  Ma, Tammy et al,  : Accelerating the path to realizing Inertial Fusion Energy via an integrated national plan,  APS DPP Meeting, Atlanta, October, 2024,
https://meetings.aps.org/Meeting/DPP24/Session/JM10.5